\documentclass[twocolumn,superscriptaddress,aps,prx]{revtex4-2}
\usepackage{amsfonts}
\usepackage{amsmath}
\usepackage{amssymb}
\usepackage{graphicx}
\usepackage{color}
\usepackage{ulem}
\usepackage{dcolumn}
\usepackage{bm}
\usepackage{url}
\usepackage[colorlinks, urlcolor = cyan, citecolor = blue, linkcolor = magenta]{hyperref}
\usepackage{ulem}
\usepackage{lipsum}
\usepackage{diagbox}
\usepackage{physics}
\def\be{\begin{equation}\begin{aligned}}
\def\ee{\end{aligned}\end{equation}}
\newcommand{\eq}[1]{\be #1 \ee}
\begin{document}

\title{Precision measurements at the interface between unitary and non-unitary encoding}
\author{Peng Xu}
\email{physicalxupeng@whu.edu.cn} 
\affiliation{Quantum Information Institute, School of Physics, Zhengzhou University, Zhengzhou 450001, China}

\begin{abstract}

We investigate precision scaling at the interface between unitary and non-unitary encoding under generalized noise including single-particle and collective dephasing and decay. Using linear response theory and the error propagation formula, we derive analytic precision expressions for both the unitary parameter $\Omega$ and the dissipation strength $\gamma$. For unitary encoding, when the observable commutes with a Hermitian noise operator, the optimal encoding time is independent of $N$, yielding the Heisenberg limit $\Delta \Omega \propto 1 / N$; otherwise the precision degrades to the standard quantum limit or ceases to improve with $N$. For non-unitary encoding, when $[\hat{A}, \hat{O}] = 0$, the precision is insensitive to intrinsic dynamics and encoding time, scaling as $\Delta \gamma \propto \sqrt{\gamma / \expval*{\hat{L}^\dagger \hat{L}}}$. Notably, for collective decay, the Dicke state reaches the Heisenberg limit $\Delta \gamma \propto 1 / N$, demonstrating that entanglement can enhance non-unitary estimation. Our results provide a unified framework and practical guidance for designing quantum metrology protocols in noisy environments.

\end{abstract}

\maketitle

\section{Introduction}
\label{sec:intro}

In quantum metrology, the precision of parameter estimation is enhanced by two key resources: quantum coherence and entanglement. Quantum coherence, arising from superposition, improves the precision scaling to $\Delta \phi \sim 1/t$, where $t$ is the encoding time~\cite{Degen2017Quantum, Braun2018Quantum}. Quantum entanglement, a nonclassical correlation shared among particles, further improves this scaling to $\Delta \phi \sim 1/N$, with $N$ the particle number~\cite{Pezz2018Quantum, Montenegro2025Quantum}. Combined, they can reach the Heisenberg limit $\Delta \phi \propto 1/(t N)$~\cite{Degen2017Quantum, Braun2018Quantum, Pezz2018Quantum, Montenegro2025Quantum, Xu2025Precision}. To compare different experiments, one normally fixes the total available time $T$, giving $\Delta \phi \propto 1 / (\sqrt{M}\, t\, N) = 1 / (\sqrt{T t}\, N)$ with $M = T / t$ the number of repeated experiments. In closed systems, these resources are independent and may be used separately or together. Quantum metrology has found applications in gravitational-wave detection~\cite{Abbott2016GW150914}, atomic clocks~\cite{Eckner2023Realizing, Cao2024Multi, Robinson2024Direct, Zaporski2025Quantum}, and electric or magnetic field sensing~\cite{Sewell2012Magnetic, Ding2022Enhanced, Mao2023Quantum, Moon2026Sensing}.

However, in realistic settings, quantum metrology is substantially impaired by environmental noise, which renders resources such as coherence and entanglement interdependent. In particular, for Markovian dephasing governed by a single-particle operator, the optimal encoding time $t$ is constrained by a decoherence time $t_c$ that scales as $t_c \propto 1 / N$~\cite{Jiao2025Quantum}. Consequently, the achievable precision degrades to $\Delta \phi \propto 1 / \sqrt{N}$, the standard quantum limit~\cite{Huelga1997Improvement, Escher2011General, Demkowicz2012Elusive}. To mitigate such noise and restore Heisenberg scaling, several external control strategies have been proposed, including quantum error correction~\cite{W2014Improved, Kessler2014Quantum, Herrera2015Quantum}, dynamical decoupling~\cite{Hahn1950Spin, Carr1954Effects, Meiboom1958Modified, Khodjasteh2005FaultTolerant, Uhrig2007Keeping, Uhrig2009Concatenated, Yao2019Uniaxial, Xu2023Floquet, Zhang2024Enhanced}, and feedback control~\cite{Goldstein2011EnvironmentAssisted, Cappellaro2012EnvironmentAssisted, Cooper2019EnvironmentAssisted, Gammelmark2014Fisher, Hirose2016Coherent, Bai2023Floquet}. Even in the absence of additional resources, precision can surpass the standard quantum limit under non-Markovian noise~\cite{Matsuzaki2011Magnetic, Chin2012Quantum, Alipour2014Quantum, Macieszczak2015Zeno, Smirne2016Ultimate}, non-phase-covariant noise~\cite{Haase2018Fundamental, Peng2024Enhanced}, or when dephasing acts perpendicular to the encoding Hamiltonian~\cite{Chaves2013Noisy, Brask2015Improved}. 

Non-Markovian noise, characterized by memory effects and information backflow, leads to non-exponential decay of coherence and entanglement, thereby enabling improved precision scaling. In particular, under non-Markovian noise, the optimal encoding time scales as $t \propto N^{- 1 / 2}$~\cite{Matsuzaki2011Magnetic, Chin2012Quantum, Alipour2014Quantum, Macieszczak2015Zeno, Smirne2016Ultimate}, yielding a precision scaling $\Delta \phi \propto N^{- 3 / 4}$ that surpasses the standard quantum limit. However, this optimal time approaches $0$ as $N \to \infty$, entering the Zeno limit, which ultimately restricts the practical performance of quantum metrology. Meanwhile, non-phase-covariant noise---which does not commute with the encoding Hamiltonian---can also improve precision scaling when the initial state is chosen appropriately. Nevertheless, such a state cannot be the optimal state for the encoding Hamiltonian; otherwise, the precision scaling would degrade under any additional perturbation, potentially reverting to the standard quantum limit~\cite{Pasquale2013Quantum, Ding2023Fundamental}. Besides, existing studies of perpendicular noise have primarily focused on single-particle dephasing~\cite{Chaves2013Noisy, Brask2015Improved}.

In this paper, we investigate precision scaling at the interface between unitary and non-unitary encoding, extending our analysis to generalized noise forms including single-particle and collective dephasing, as well as decay. We first determine the relation between the optimal encoding time $t$ and the particle number $N$, and then derive the precision scaling $\Delta \Omega$ for unitary encoding under these noise types, assuming the initial state is optimal for the encoding Hamiltonian. By combining linear response theory with the error propagation formula, we show that when the measured observable commutes with the noise operator, the optimal encoding time $t$ becomes independent of $N$, yielding $\Delta \Omega \propto 1 / \expval*{\Delta \hat{A}} \propto 1 / N$. We also analyze the precision scaling $\Delta \gamma$ for non-unitary encoding, where $\gamma$ is encoded through the noise operator. When the observable commutes with the unitary part, the precision scales as $\Delta \gamma \propto \sqrt{\gamma / \expval*{\hat{L}^\dagger \hat{L}}}$. Our results provide a unified perspective on precision scaling at the interface between unitary and non-unitary encoding, and offer guidance for designing quantum metrology protocols in noisy environments.

\section{Unitary encoding under noise}
\label{sec:uni_pre}

We consider a system undergoing unitary encoding by an external field while simultaneously subject to noise. The system dynamics are governed by the following non-Hermitian Hamiltonian~\cite{Scully1997Quantum, Pan2020Non, Xu2025Precision}
\eq{
    \hat{H} = \hat{H}_0 + \hat{H}_{\text{d}},
    \label{eq:hamil}
}
where 
\eq{
    \hat{H}_0 &= \Omega \hat{A}, \\
    \hat{H}_{\text{d}} &= - i \gamma \hat{L}^{\dagger} \hat{L} + [\hat{\xi}(t) \hat{L}^\dagger + \hat{\xi}^\dagger(t) \hat{L}],
}
with $\hat{A}$ the encoding Hamiltonian and $\hat{L}$ the noise operator. The noise field $\hat{\xi}(t)$ satisfies $\expval*{\hat{\xi}(t)} = 0$ and $\expval*{\hat{\xi}(t) \hat{\xi}^\dagger(t')} = 2 \gamma \delta(t - t')$. The parameter $\Omega$ denotes the encoding strength, while $\gamma$ characterizes the noise strength.

For an observable $\hat{O}$, its Heisenberg-picture expectation value can be expressed in terms of the interaction-picture operator as
\eq{
    \expval*{\hat{O}_\text{H}(t)} = \expval*{\hat{U}^\dagger(t) \hat{O}_\text{I}(t) \hat{U}(t)},
    \label{eq:picture}
}
where $\hat{U}(t) = \mathcal{T} e^{- i \int_0^t \hat{H}_{0, \text{I}}(\tau) \dd \tau}$, and $\hat{O}_{\text{I}}(t)$ is the interaction-picture operator with respect to the dissipative part, given by $\hat{O}_{\text{I}}(t)=\mathcal{T} e^{i \int_0^t \hat{H}_{\text{d}}^\dagger(\tau) \dd \tau} \hat{O} \mathcal{T} e^{-i \int_0^t \hat{H}_{\text{d}}(\tau) \dd \tau}$. For small $\Omega$, expanding $\hat{U}(t)$ to first order yields
\eq{
    \delta \expval*{\hat{O}(t)} = - i \Omega \int_0^t \expval*{[\hat{O}_\text{I}(t), \hat{A}_\text{I}(\tau)]} \dd \tau.
}
The error propagation formula~\cite{note1} then gives the estimation precision for $\Omega$:
\eq{
    \Delta \Omega = \frac{\Delta \hat{O}_{\text{H}}(t)}{\sqrt{M} \abs*{\int_0^t \expval*{[\hat{O}_\text{I}(t), \hat{A}_\text{I}(\tau)]} \dd \tau}}.
    \label{eq:uni_prec}
}

To determine the optimal encoding time $t$, we analyze the behavior of the numerator and denominator in Eq.~\eqref{eq:uni_prec}. The numerator $\Delta \hat{O}_{\text{H}}(t)$ is the Heisenberg-picture fluctuation of $\hat{O}$; it is amplified by the noise and may grow with time for the initially optimal state. The denominator $\abs*{\int_0^t \expval*{[\hat{O}_\text{I}(t), \hat{A}_\text{I}(\tau)]} \dd \tau}$ quantifies the response of $\hat{O}$ to variations in $\Omega$, and typically grows over time owing to the accumulation of phase information. Noise therefore introduces a trade-off: longer encoding times enhance the response but also increase the fluctuations. By analyzing this trade-off, we identify the optimal $t$ that minimizes $\Delta \Omega$, and subsequently determine the precision scaling with respect to the particle number $N$.

Both $\hat{O}_{\text{I}}(t)$ and $\hat{A}_{\text{I}}(\tau)$ reside in the interaction picture and are influenced by the noise operator $\hat{L}$. Applying the cumulant expansion, the denominator expands to first order in $\gamma$ as (see Appendix~\ref{sec:app_deri} for details)
\eq{
    &\int_0^t \expval*{[\hat{O}_\text{I}(t), \hat{A}_\text{I}(\tau)]} \dd \tau = t \expval*{[\hat{O}, \hat{A}]} \\
    &+ \frac{3}{2} \gamma t^2 (\expval*{[\hat{L}^\dagger, \hat{O}]} \expval*{[\hat{L}, \hat{A}]} + \expval*{[\hat{O}, \hat{L}]} \expval*{[\hat{L}^\dagger, \hat{A}]}).
    \label{eq:uni_deno}
}
Similarly, applying the cumulant expansion to the numerator yields, to first order in $\gamma$ (see Appendix~\ref{sec:app_deri} for details),
\eq{
    &\Delta \hat{O}_{\text{H}}^2(t) = \expval*{\Delta \hat{O}^2} \\
    &+ 2 \gamma t (\expval*{\hat{L}^\dagger \hat{O}}_{\text{c}} \expval*{[\hat{O}, \hat{L}]} + \expval*{\hat{O} \hat{L}}_{\text{c}} \expval*{[\hat{L}^\dagger, \hat{O}]}),
    \label{eq:uni_nume}
}
where $\expval*{\hat{L}^\dagger \hat{O}}_{\text{c}} = \expval*{\hat{L}^\dagger \hat{O}} - \expval*{\hat{L}^\dagger} \expval*{\hat{O}}$ and $\expval*{\hat{O} \hat{L}}_{\text{c}} = \expval*{\hat{O} \hat{L}} - \expval*{\hat{O}} \expval*{\hat{L}}$. The precision for estimating $\Omega$ for unitary encoding under noise can thus be expressed as

{\footnotesize\eq{
    \Delta \Omega = \frac{\sqrt{\expval*{\Delta \hat{O}}^2 + 2 \gamma t (\expval*{\hat{L}^\dagger \hat{O}}_{\text{c}} \expval*{[\hat{O}, \hat{L}]} + \expval*{\hat{O} \hat{L}}_{\text{c}} \expval*{[\hat{L}^\dagger, \hat{O}]})}}{\abs*{\expval*{[\hat{O}, \hat{A}]} + \frac{3}{2} \gamma t (\expval*{[\hat{L}^\dagger, \hat{A}]} \expval*{[\hat{O}, \hat{L}]} +  \expval*{[\hat{L}, \hat{A}]}  \expval*{[\hat{L}^\dagger, \hat{O}]} )} \sqrt{T t}}.
    \label{eq:uni_prec_final}
}}

From Eq.~\eqref{eq:uni_prec_final}, if $\hat{L}$ is Hermitian ($\hat{L}^\dagger = \hat{L}$) and $\hat{O}$ commutes with $\hat{L}$ ($\expval*{[\hat{O}, \hat{L}]} = 0$), the precision simplifies to
\eq{
    \Delta \Omega = \frac{\expval*{\Delta \hat{O}}}{\abs*{\expval*{[\hat{O}, \hat{A}]}} \sqrt{T t}}.
    \label{eq:uni_prec_final_1}
}
In this case, the precision is independent of the noise strength $\gamma$. The optimal encoding time $t$ may therefore be chosen arbitrarily before the steady state is reached, yielding the scaling $\Delta \Omega \propto 1 / \expval*{\Delta \hat{A}} \propto 1 / N$ for an initially optimal probe state~\cite{Xu2025Precision}. If $\hat{O}$ does not commute with $\hat{L}$, the precision becomes sensitive to $\gamma$, and the optimal $t$ may depend on $N$, leading to a modified scaling. For a non-Hermitian noise operator ($\hat{L}^\dagger \neq \hat{L}$), Eq.~\eqref{eq:uni_prec_final} admits no such simplification; the $N$-dependence of the optimal encoding time then depends on the specific forms of $\hat{A}$, $\hat{O}$, $\hat{L}$, and on the initial state of the system.

\subsection{Dephasing noise}
\label{sec:uni_dephasing}

\begin{figure}[t!]
    \centering
    \includegraphics[width=8.5cm]{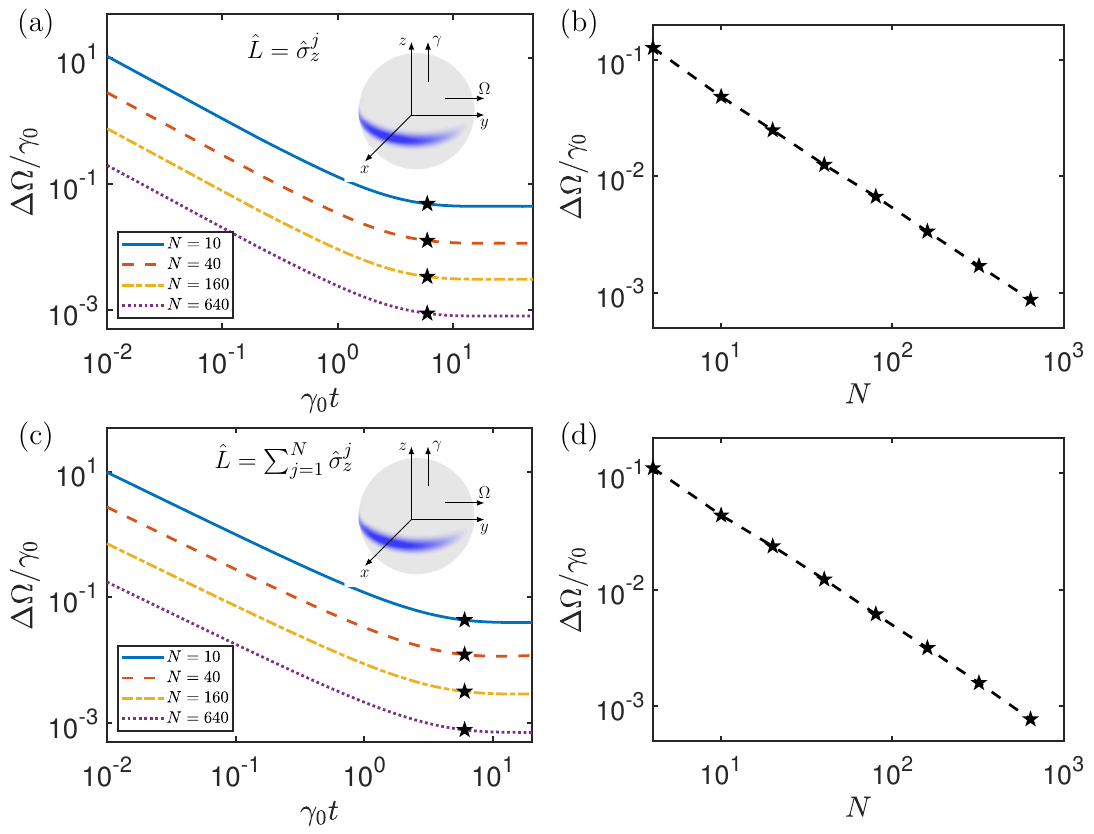}
    \caption{Precision dynamics for various $N$ under single-particle dephasing (a) and collective dephasing (c). The initial state is the optimal two-axis twisted state with $\expval*{\hat{J}_x} \propto N$ and $\expval*{\Delta \hat{J}_z} \propto 1$. (a, c) The optimal encoding time $t$, indicated by pentagrams at the maximum curvature, is independent of $N$. (b, d) The precision scaling at the optimal time exhibits the Heisenberg scaling $\Delta \Omega \propto 1 / N$. The dissipation rate is $\gamma / \gamma_0 = 0.1$. The energy and time units are $\gamma_0$ and $\gamma_0^{- 1}$, respectively.}
    \label{fig:uni_dephasing_1}
\end{figure}

We first consider Hermitian $\hat{L}$, corresponding to dephasing noise. Depending on the spatial correlation of the noise, this case splits into two main categories: single-particle dephasing ($\hat{L} = \hat{\sigma}_z^j$) and collective dephasing ($\hat{L} = \hat{J}_z = \sum_{j = 1}^N \hat{\sigma}_z^j$). To satisfy $\expval*{[\hat{O}, \hat{L}]} = 0$, the observable must be $\hat{O} = \hat{J}_z = \sum_{j = 1}^N \hat{\sigma}_z^j$; without loss of generality, we choose the encoding operator $\hat{A} = \hat{J}_y = \sum_{j = 1}^N \hat{\sigma}_y^j$. The precision for estimating $\Omega$ then becomes
\eq{
    \Delta \Omega = \frac{\expval*{\Delta \hat{J}_z}}{\abs*{2 \expval*{\hat{J}_x}} \sqrt{T t}}.
    \label{eq:uni_prec_dephasing}
}
For a coherent spin state $\bigotimes_{j=1}^N \ket{+}_j$, the eigenstate of $\hat{J}_x$ with eigenvalue $N$, we have $\expval*{\Delta \hat{J}_z} = \sqrt{N}$ and $\expval*{\hat{J}_x} = N$. The precision therefore scales as $\Delta \Omega \propto 1 / (\sqrt{N T t})$, coinciding with the standard quantum limit -- as expected for an unentangled initial state. By contrast, choosing the optimal two-axis twisted state, with $\expval*{\hat{J}_x} \propto N$ and $\expval*{\Delta \hat{J}_z} \propto 1$, the precision reaches the Heisenberg limit $\Delta \Omega \propto 1 / (N \sqrt{T t})$.

\begin{figure}[t!]
    \centering
    \includegraphics[width=8.5cm]{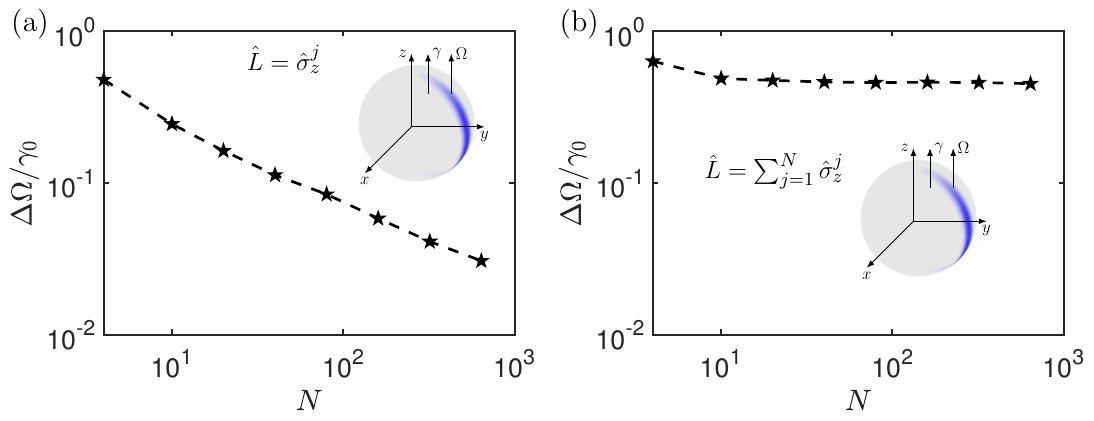}
    \caption{Precision scaling at the optimal encoding time for various $N$ under single-particle dephasing (a) and collective dephasing (b). The initial state is the optimal two-axis twisted state with $\expval*{\hat{J}_y} \propto N$ and $\expval*{\Delta \hat{J}_x} \propto 1$. The precision scaling degrades to $\Delta \Omega \propto 1 / \sqrt{N}$ under single-particle dephasing and $\Delta \Omega \propto 1$ under collective dephasing. The dissipation rate is $\gamma / \gamma_0 = 0.1$.}
    \label{fig:uni_dephasing_2}
\end{figure}

To validate these results, we numerically calculate $\Delta \Omega$ using the error propagation formula together with the discrete truncated Wigner approximation (DTWA) method~\cite{note1, Yoneya2025Path}. We set $M = 1$ to determine the optimal encoding time for a single experimental realization. As shown in Fig.~\ref{fig:uni_dephasing_1}, the optimal $t$ is identified from the point of maximum curvature in the precision dynamics, confirming its independence of $N$---see Fig.~\ref{fig:uni_dephasing_1}(a) for single-particle dephasing and Fig.~\ref{fig:uni_dephasing_1}(c) for collective dephasing. Although Eq.~\eqref{eq:uni_prec_dephasing} predicts that the precision decreases monotonically with $t$, the numerics exhibit a plateau beyond the optimal time because the system approaches a steady state as $t \sim 1 / \gamma$. The precision scaling at the optimal time [Figs.~\ref{fig:uni_dephasing_1}(b) and \ref{fig:uni_dephasing_1}(d)] demonstrates the Heisenberg scaling $\Delta \Omega \propto 1 / N$. Intuitively, the Heisenberg limit is expected here because the dephasing noise does not affect the fluctuations of $\hat{J}_z$.  

When $\hat{O}$ does not commute with $\hat{L}$, we choose, without loss of generality, $\hat{O} = \hat{J}_x$ and $\hat{A} = \hat{J}_z$. The precision for estimating $\Omega$ then becomes
\eq{
    \Delta \Omega = \frac{\sqrt{\expval*{\Delta \hat{J}_x}^2 + 8 \gamma t \sum_j \expval*{\hat{\sigma}_y^j}^2}}{\abs{2 \expval*{\hat{J}_y}} \sqrt{T t}},
    \label{eq:uni_prec_dephasing_2}
}
for single-particle dephasing, and
\eq{
    \Delta \Omega = \frac{\sqrt{\expval*{\Delta \hat{J}_x}^2 + 8 \gamma t \expval*{\hat{J}_y}^2}}{\abs{2 \expval*{\hat{J}_y}} \sqrt{T t}},
    \label{eq:uni_prec_dephasing_3}
}
for collective dephasing. When the initial state is the optimal two-axis twisted state, with $\expval*{\hat{J}_y} \propto N$ and $\expval*{\Delta \hat{J}_x} \propto 1$, the precision at the optimal time, as shown in Fig.~\ref{fig:uni_dephasing_2}, degrades to $\Delta \Omega \propto 1 / \sqrt{N}$ for single-particle dephasing and $\Delta \Omega \propto 1$ for collective dephasing. Intuitively, single-particle dephasing introduces additional fluctuations in $\hat{J}_x$ scaling as $N$, leading to $t \propto 1 / N$ and degrading the precision scaling to the standard quantum limit. Collective dephasing, meanwhile, introduces fluctuation terms scaling as $N^2$, resulting in $t \propto 1 / N^2$, which dominates the fluctuations of $\hat{J}_x$ and prevents any improvement in precision with $N$. When the initial state is the eigenstate of $\hat{J}_y$ with eigenvalue $N$, $\expval*{\hat{J}_y} = N$ and $\expval*{\Delta \hat{J}_x} = \sqrt{N}$, the precision scales as $1 / \sqrt{N}$ under both types of dephasing. Thus, when $[\hat{O}, \hat{L}] \neq 0$, a product state can exhibit better precision scaling than an entangled state. An additional case with $[\hat{O}, \hat{L}] \neq 0$ ( $\hat{O} = \hat{J}_x$ and $\hat{A} = \hat{J}_y$), is discussed in Appendix~\ref{sec:app_uni_dephasing}.

\subsection{Decay noise}
\label{sec:uni_decay}

\begin{figure}[t!]
    \centering
    \includegraphics[width=8.5cm]{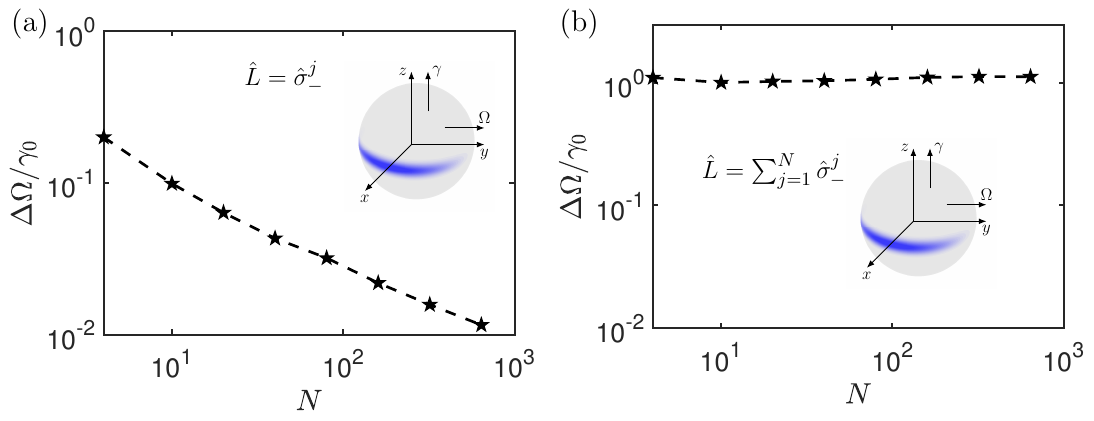}
    \caption{Precision scaling at the optimal encoding time for various $N$ under single-particle decay (a) and collective decay (b). The initial state is the optimal two-axis twisted state with $\expval*{\hat{J}_x} \propto N$ and $\expval*{\Delta \hat{J}_z} \propto 1$. The precision scaling degrades to $\Delta \Omega \propto 1 / \sqrt{N}$ under single-particle decay and $\Delta \Omega \propto 1$ under collective decay. The dissipation rate is $\gamma / \gamma_0 = 0.1$.}
    \label{fig:uni_decay_1}
\end{figure}

Next, we consider non-Hermitian $\hat{L}$, corresponding to decay noise: $\hat{L} = \hat{\sigma}_-^j$ for single-particle decay and $\hat{L} = \hat{J}_- = \sum_{j = 1}^N \hat{\sigma}_-^j$ for collective decay. The observable and encoding operator are chosen analogously to the dephasing case; however, since $\hat{L}$ is non-Hermitian, $\hat{O}$ cannot commute with it. Without loss of generality, we take $\hat{O} = \hat{J}_z$ and $\hat{A} = \hat{J}_y$, with the initial state being the optimal two-axis twisted state ($\expval*{\hat{J}_x} \propto N$, $\expval*{\Delta \hat{J}_z} \propto 1$). The precision $\Delta \Omega$ is shown in Fig.~\ref{fig:uni_decay_1}: it degrades to $\Delta \Omega \propto 1 / \sqrt{N}$ for single-particle decay and $\Delta \Omega \propto 1$ for collective decay, due to different optimal encoding time scales. Overall, the optimal encoding time is proportional to $1 / \expval*{\hat{L}^\dagger \hat{L}}$ based on the flucuation-dissipation theorem, for both decoherence processes of dephasing and decay noise. Specifically, it scales as $1/N$ for single-particle decay and $1/N^2$ for collective decay, leading to the standard quantum limit and no improvement with $N$, respectively.

\section{Non-unitary encoding with intrinsic dynamics}
\label{sec:nonuni_pre}

Although estimating a signal parameter is central to quantum metrology, estimating the dissipation strength $\gamma$ is also of practical use---for instance, to identify both the type and the strength of dissipative channels. However, intrinsic dynamics---such as Larmor precession and interaction scattering---still affect the evolution when estimating $\gamma$. Hence, the system remains described by Eq.~\eqref{eq:hamil}. We follow the same procedure as in the unitary encoding case, with the key difference that $\gamma$ is now encoded through the non-unitary noise operator $\hat{L}$. Accordingly, the Heisenberg--interaction-picture relation for $\hat{O}$ is built via $\hat{U} = \mathcal{T} e^{- i \int_0^t \hat{H}_{\text{d}, \text{I}}(\tau) \dd \tau}$, with the interaction-picture operator given by $\hat{O}_\text{I}(t) = \mathcal{T} e^{i \int_0^t \hat{H}_0^\dagger(\tau) \dd \tau} \hat{O} \mathcal{T} e^{- i \int_0^t \hat{H}_0(\tau) \dd \tau}$. For small $\gamma$, expanding $\hat{U}(t)$ to first order and substituting into Eq.~\eqref{eq:picture} yields
\eq{
    \delta \expval*{\hat{O}(t)} &= \gamma \int_0^t [ 2 \expval*{\hat{L}_\text{I}^\dagger(\tau) \hat{O}_\text{I}(t) \hat{L}_\text{I}(\tau)} \\
    &\quad - \expval*{\{\hat{L}_\text{I}^\dagger(\tau) \hat{L}_\text{I}(\tau), \hat{O}_\text{I}(t)\}}] \dd \tau,
}
which is the non-Hermitian linear response theory~\cite{Pan2020Non}. The error propagation formula then gives the estimation precision for $\gamma$:

{\footnotesize\eq{
    \Delta \gamma = \frac{\Delta \hat{O}_\text{H}(t)}{ \sqrt{M} \abs*{ \int_0^t [ 2 \expval*{\hat{L}_\text{I}^\dagger(\tau) \hat{O}_\text{I}(t) \hat{L}_\text{I}(\tau)} - \expval*{\{\hat{L}_\text{I}^\dagger(\tau) \hat{L}_\text{I}(\tau), \hat{O}_\text{I}(t)\}}] \dd \tau}}. 
}}

As shown in our previous work~\cite{Xu2025Precision}, in the absence of intrinsic dynamics, the optimal precision for estimating $\gamma$ is attained when the initial state is an eigenstate of $\hat{O}$. To investigate how intrinsic dynamics affects the precision of non-unitary encoding---in analogy with the unitary encoding case, where the optimal squeezed state was chosen---we likewise take the initial state to be an eigenstate of $\hat{O}$. Under this condition, applying the cumulant expansion expands the denominator to first order in $\Omega$ as (see Appendix~\ref{sec:app_deri_nonuni} for details)
\eq{
    &\int_0^t [ 2 \expval*{\hat{L}_\text{I}^\dagger(\tau) \hat{O}_\text{I}(t) \hat{L}_\text{I}(\tau)} - \expval*{\{\hat{L}_\text{I}^\dagger(\tau) \hat{L}_\text{I}(\tau), \hat{O}_\text{I}(t)\}}] \dd \tau \\
    &= t [2 \expval*{\hat{L}^\dagger \hat{O} \hat{L}} - \expval*{\{\hat{L}^\dagger \hat{L}, \hat{O}\}} ] \\
    &\quad + i \Omega t^2 \expval*{\hat{L}^\dagger} (\expval*{[\hat{A}, \hat{O}] \hat{L}} + \expval*{\hat{L} [\hat{O}, \hat{A}]}) \\
    &\quad + i \Omega t^2 \expval*{\hat{L}} (\expval*{[\hat{O}, \hat{A}] \hat{L}^\dagger} + \expval*{\hat{L}^\dagger [\hat{A}, \hat{O}]}).
    \label{eq:nonuni_deno}
}
Similarly, the numerator expands to (see Appendix~\ref{sec:app_deri_nonuni} for details)
\eq{
    \Delta \hat{O}_{\text{H}}^2(t) = 2 \gamma t \expval*{[\hat{L}^\dagger, \hat{O}][\hat{O}, \hat{L}]}.
    \label{eq:nonuni_nume}
}

Furthermore, when $[\hat{A}, \hat{O}] = 0$, the precision simplifies to
\eq{
    \Delta \gamma = \frac{\sqrt{2 \gamma \expval*{[\hat{L}^\dagger, \hat{O}][\hat{O}, \hat{L}]}}}{\abs*{2 \expval*{\hat{L}^\dagger \hat{O} \hat{L}} - \expval*{\{\hat{L}^\dagger \hat{L}, \hat{O}\}}} \sqrt{T}}
}
Notably, the precision is independent of intrinsic dynamics. More importantly, it is independent of $t$; this implies that even when $[\hat{A}, \hat{O}] \neq 0$, the precision scaling remains unaffected by intrinsic dynamics until the steady state is reached. Finally, the expression further reduces to $\Delta \gamma \propto \sqrt{\gamma / \expval*{\hat{L}^\dagger \hat{L}}}$~\cite{Xu2025Precision}.

\subsection{Dephasing processes}
\label{sec:nonuni_dephasing}

\begin{figure}[t!]
    \centering
    \includegraphics[width=8.5cm]{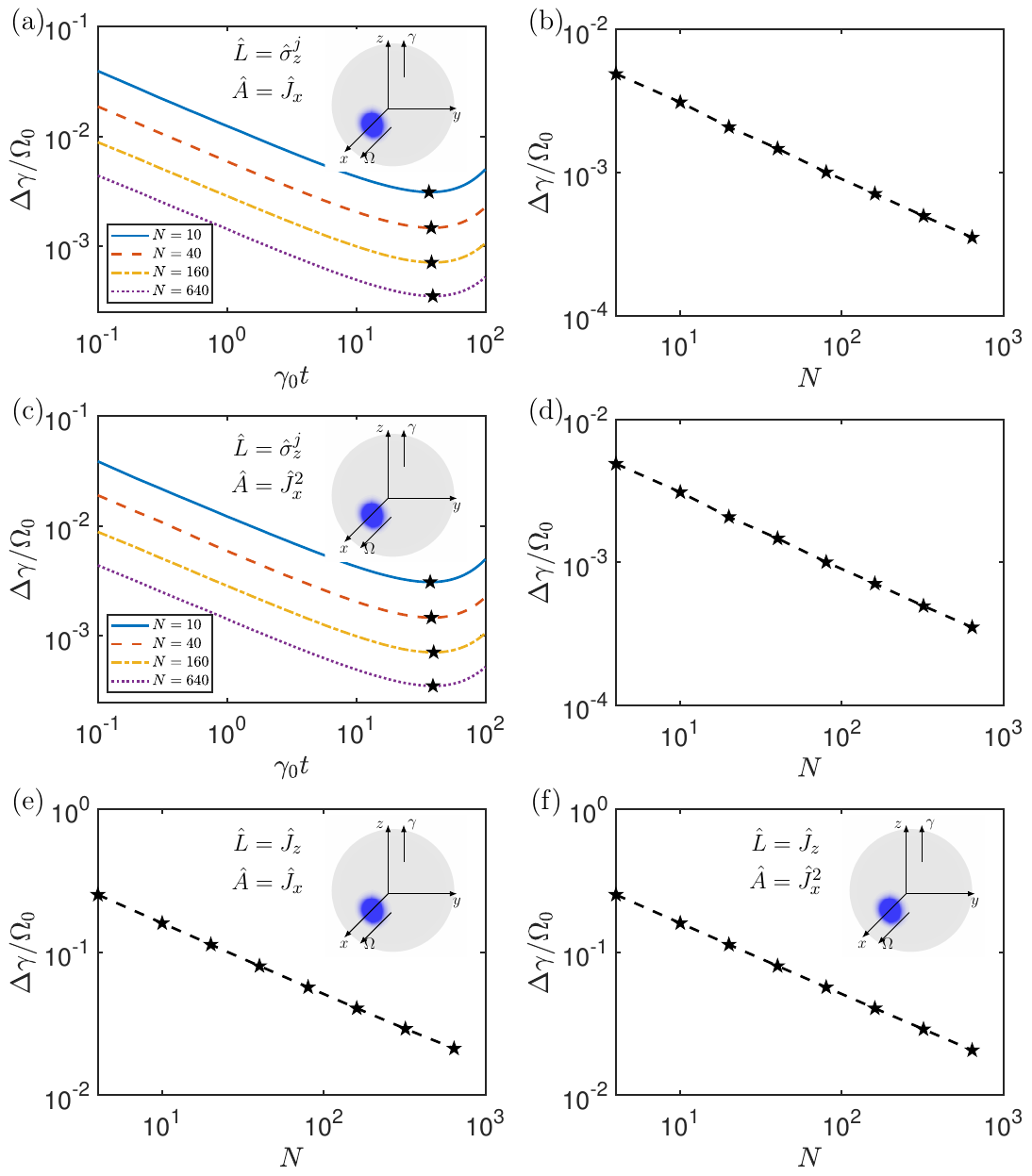}
    \caption{Precision dynamics for various $N$ under non-unitary encoding with $\hat{L} = \hat{\sigma}_z^j$, $\hat{O} = \hat{J}_x$, and intrinsic dynamics $\hat{A} = \hat{J}_x$ (a) and $\hat{A} = \hat{J}_x^2$ (c). The initial state is the eigenstate of $\hat{J}_x$ with eigenvalue $N$. The precision scaling at the optimal time, indicated by pentagrams at the maximum curvature, yields $\Delta \gamma \propto 1 / \sqrt{N}$ for both cases [panels (b) and (d)]. Panels (e) and (f) show the same scaling $\Delta \gamma \propto 1 / \sqrt{N}$ for $\hat{L} = \hat{J}_z$. The intrinsic-dynamics strength is $\Omega / \Omega_0 = 0.1$. The energy and time units are $\Omega_0$ and $\Omega_0^{- 1}$, respectively.}
    \label{fig:nonuni_dephasing_1}
\end{figure}

We first consider dephasing processes, with $\hat{L} = \hat{\sigma}_z^j$ (single-particle) or $\hat{L} = \hat{J}_z$ (collective). To satisfy $[\hat{A}, \hat{O}] = 0$, we choose $\hat{O} = \hat{J}_x$ and either $\hat{A} = \hat{J}_x$ or $\hat{A} = \hat{J}_x^2$. The precision for estimating $\gamma$ then becomes
\eq{
    \Delta \gamma = \frac{\sqrt{2 \gamma N}}{\abs*{2 \sum_j \expval*{\hat{\sigma}_x^j}} \sqrt{T}}.
}
for single-particle dephasing, and
\eq{
    \Delta \gamma = \frac{\sqrt{2 \gamma \expval*{\Delta \hat{J}_y}^2}}{\abs*{2 \expval*{\hat{J}_x}} \sqrt{T}}.
}
for collective dephasing. Choosing the eigenstate of $\hat{J}_x$ with eigenvalue $N$ as the initial state, the precision scales as $\Delta \gamma \propto 1 / \sqrt{N}$ in both cases. Numerical results from the error propagation formula with the DTWA method confirm this scaling (Fig.~\ref{fig:nonuni_dephasing_1}). The precision at the optimal time is independent of the intrinsic dynamics---whether Larmor precession ($\hat{A} = \hat{J}_x$) or interaction scattering ($\hat{A} = \hat{J}_x^2$). Moreover, the scaling is independent of the noise correlation, as illustrated in Fig.~\ref{fig:nonuni_dephasing_1}(e) and Fig.~\ref{fig:nonuni_dephasing_1}(f) for collective dephasing. These conclusions are consistent with the analytical expression, which shows that the precision is independent of both intrinsic dynamics and noise correlation whenever $[\hat{A}, \hat{O}] = 0$. For a Dicke state $\ket{N, M}_x$ with an eigenvalue $M$ of $\hat{J}_x$ smaller than $N$, the precision further degrades because $\expval*{\Delta \hat{J}_y}$ increases and $\expval*{\hat{J}_x}$ decreases. The case $[\hat{A}, \hat{O}] \neq 0$ is treated in Appendix~\ref{sec:app_nonuni_dephasing}.

\subsection{Decay processes}
\label{sec:nonuni_decay}

\begin{figure}[t!]
    \centering
    \includegraphics[width=8.5cm]{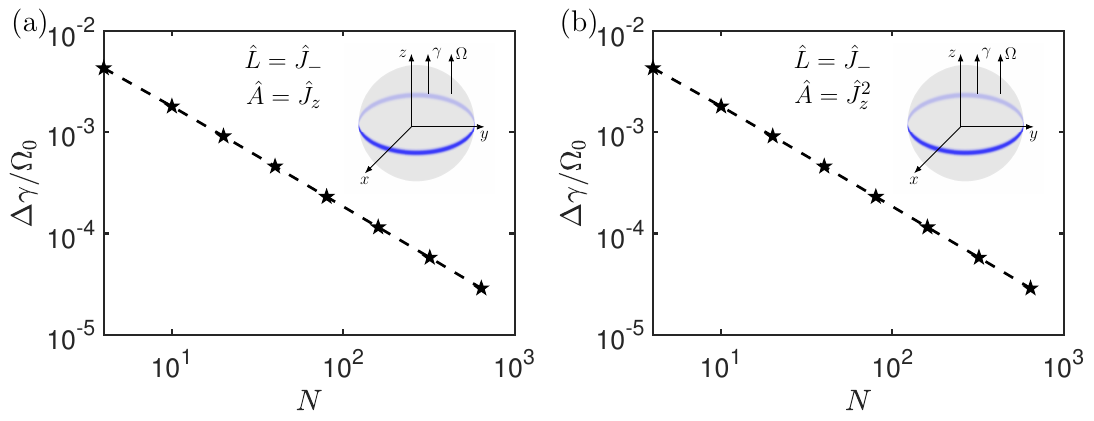}
    \caption{Precision scaling at the optimal encoding time for different total particle number $N$ under non-unitary encoding $\hat{L} = \hat{J}_-$ with intrinsic dynamics $\hat{A} = \hat{J}_z$ (a) or $\hat{A} = \hat{J}_z^2$ (b), where the observable is $\hat{O} = \hat{J}_z$. The initial state is the Dicke state $\ket{N, 0}_z$ with eigenvalue $0$ of $\hat{J}_z$. The precision scales as the Heisenberg limit $\Delta \gamma \propto 1 / N$. $\Omega / \Omega_0 = 0.1$ for the intrinsic dynamics. The energy and the time units are $\Omega_0$ and $\Omega_0^{- 1}$, respectively.}
    \label{fig:nonuni_decay_1}
\end{figure}

Next, we consider decay processes, with $\hat{L} = \hat{\sigma}_-^j$ (single-particle) or $\hat{L} = \hat{J}_-$ (collective). Without loss of generality, we choose $\hat{O} = \hat{J}_z$ and $\hat{A} = \hat{J}_z$ or $\hat{A} = \hat{J}_z^2$, such that $[\hat{A}, \hat{O}] = 0$. The precision for estimating $\gamma$ then becomes
\eq{
    \Delta \gamma = \frac{\sqrt{2 \gamma N}}{\abs*{\sum_j \expval*{\hat{\sigma}_z^j} + N} \sqrt{T}},
}
for single-particle decay, and
\eq{
    \Delta \gamma = \frac{\sqrt{2 \gamma \expval*{\hat{J}_+ \hat{J}_-}}}{\abs*{2 \expval*{\hat{J}_+ \hat{J}_-}} \sqrt{T}},
}
for collective decay. From the single-particle expression, the precision scales as $\Delta \gamma \propto 1 / \sqrt{N}$ regardless of whether the initial state is a product state or an entangled state, provided it is an eigenstate of $\hat{J}_z$. Notably, in contrast to dephasing, the precision for decay processes is dependent of the noise correlation. From the collective expression, the precision scales as $\Delta \gamma \propto 1 / \sqrt{N}$ for the polarized state $\ket{N, N}_z$ but reaches $\Delta \gamma \propto 1 / N$ for the Dicke state $\ket{N, 0}_z$. Numerical results for the Dicke state, as shown in Fig.~\ref{fig:nonuni_decay_1}, confirm the Heisenberg scaling $\Delta \gamma \propto 1 / N$ at the optimal time.

\section{Conclusions}

We have investigated precision scaling at the interface between unitary and non-unitary encoding under generalized noise models including single-particle and collective dephasing and decay. By combining linear response theory with the error propagation formula, we derived explicit precision expressions for both the unitary parameter $\Omega$ and the dissipation strength $\gamma$. For unitary encoding, when the observable commutes with a Hermitian noise operator, the optimal encoding time is independent of $N$ and the precision reaches the Heisenberg limit $\Delta \Omega \propto 1 / N$. Otherwise, the precision degrades to the standard quantum limit under single-particle noise, or ceases to improve with $N$ under collective noise. For non-unitary encoding, when $[\hat{A}, \hat{O}] = 0$, the precision is independent of intrinsic dynamics and encoding time, scaling as $\Delta \gamma \propto \sqrt{\gamma / \expval*{\hat{L}^\dagger \hat{L}}}$. In collective decay, the Dicke state achieves the Heisenberg limit $\Delta \gamma \propto 1 / N$, demonstrating that entanglement can enhance non-unitary estimation. Our results provide a unified perspective and practical guidance for quantum metrology in noisy environments.

{\bf Acknowledgements} This project is supported by the National Natural Science Foundation of China (Grant No. 12375023 and No. 12204428); the Natural Science Foundation of Henan Province (Grant No.242300421159).

\appendix

\widetext{

\section{Derivations for Eqs.~\eqref{eq:uni_deno} and~\eqref{eq:uni_nume}}
\label{sec:app_deri}

Both $\hat{O}_{\text{I}}(t)$ and $\hat{A}_{\text{I}}(t)$ are influenced by $\hat{L}$ and expand to first order in $\gamma$ as
\eq{
    \hat{O}_\text{I}(t) &= \hat{O} + \gamma t [2 \hat{L}^\dagger \hat{O} \hat{L} - \hat{L}^\dagger \hat{L} \hat{O} - \hat{O} \hat{L}^\dagger \hat{L}], \\
    \hat{A}_\text{I}(t) &= \hat{A} + \gamma t [2 \hat{L}^\dagger \hat{A} \hat{L} - \hat{L}^\dagger \hat{L} \hat{A} - \hat{A} \hat{L}^\dagger \hat{L}].
}
Substituting into Eq.~\eqref{eq:uni_prec} and retaining terms to first order in $\gamma$ gives
\eq{
    \int_0^t \expval*{[\hat{O}_\text{I}(t), \hat{A}_\text{I}(\tau)]} \dd \tau &= t \expval*{[\hat{O}, \hat{A}]} + \gamma t^2 \expval*{[2 \hat{L}^\dagger \hat{O} \hat{L} - \hat{L}^\dagger \hat{L} \hat{O} - \hat{O} \hat{L}^\dagger \hat{L}, \hat{A}]} + \frac{1}{2} \gamma t^2 \expval*{[\hat{O}, 2 \hat{L}^\dagger \hat{A} \hat{L} - \hat{L}^\dagger \hat{L} \hat{A} - \hat{A} \hat{L}^\dagger \hat{L}]}.
}
Applying the cumulant expansion to two-point correlations simplifies each term as
\eq{
    2 \expval*{[\hat{L}^\dagger \hat{O} \hat{L}, \hat{A}]} &= 2 \expval*{\hat{L}^\dagger \hat{O}} \expval*{[\hat{L}, \hat{A}]} + 2 \expval*{\hat{O} \hat{L}} \expval*{[\hat{L}^\dagger, \hat{A}]} + 2 \expval*{\hat{L}^\dagger \hat{L}} \expval*{[\hat{O}, \hat{A}]}, \\
    - \expval*{[\hat{L}^\dagger \hat{L} \hat{O}, \hat{A}]} &= - \expval*{\hat{L}^\dagger \hat{O}} \expval*{[\hat{L}, \hat{A}]} - \expval*{\hat{L} \hat{O}} \expval*{[\hat{L}^\dagger, \hat{A}]} - \expval*{\hat{L}^\dagger \hat{L}} \expval*{[\hat{O}, \hat{A}]}, \\
    - \expval*{[\hat{O} \hat{L}^\dagger \hat{L}, \hat{A}]} &= - \expval*{\hat{O} \hat{L}^\dagger} \expval*{[\hat{L}, \hat{A}]} - \expval*{\hat{O} \hat{L}} \expval*{[\hat{L}^\dagger, \hat{A}]} - \expval*{\hat{L}^\dagger \hat{L}} \expval*{[\hat{O}, \hat{A}]}.
}
Hence
\eq{
    \expval*{[2 \hat{L}^\dagger \hat{O} \hat{L} - \hat{L}^\dagger \hat{L} \hat{O} - \hat{O} \hat{L}^\dagger \hat{L}, \hat{A}]} &= \expval*{[\hat{L}^\dagger, \hat{O}]} \expval*{[\hat{L}, \hat{A}]} + \expval*{[\hat{O}, \hat{L}]} \expval*{[\hat{L}^\dagger, \hat{A}]}.
}
A similar calculation yields
\eq{
    \expval*{[\hat{O}, 2 \hat{L}^\dagger \hat{A} \hat{L} - \hat{L}^\dagger \hat{L} \hat{A} - \hat{A} \hat{L}^\dagger \hat{L}]} &= \expval*{[\hat{L}^\dagger, \hat{O}]} \expval*{[\hat{L}, \hat{A}]} + \expval*{[\hat{O}, \hat{L}]} \expval*{[\hat{L}^\dagger, \hat{A}]}.
}
The denominator thus reduces to
\eq{
    \int_0^t \expval*{[\hat{O}_\text{I}(t), \hat{A}_\text{I}(\tau)]} \dd \tau = t \expval*{[\hat{O}, \hat{A}]} + \frac{3}{2} \gamma t^2 \{\expval*{[\hat{L}^\dagger, \hat{O}]} \expval*{[\hat{L}, \hat{A}]} + \expval*{[\hat{O}, \hat{L}]} \expval*{[\hat{L}^\dagger, \hat{A}]}\},
}
which is Eq.~\eqref{eq:uni_deno}.

The numerator $\Delta \hat{O}_{\text{H}}^2(t)$ expands to first order in $\gamma$ and $\Omega$ as
\eq{
    \Delta \hat{O}_{\text{H}}^2(t) &= [\expval*{\hat{O}^2} - \expval*{\hat{O}}^2] + i \Omega t \{\expval*{[\hat{A}, \hat{O}^2]} - 2 \expval*{\hat{O}} \expval*{[\hat{A}, \hat{O}]}\} \\
    &\quad + \gamma t \{2 \expval*{\hat{L}^\dagger \hat{O}^2 \hat{L}} - \expval*{\hat{L}^\dagger \hat{L} \hat{O}^2} - \expval*{\hat{O}^2 \hat{L}^\dagger \hat{L}}\} - 2 \gamma t \{2 \expval*{\hat{O}} \expval*{\hat{L}^\dagger \hat{O} \hat{L}} - \expval*{\hat{O}} \expval*{\hat{L}^\dagger \hat{L} \hat{O}} - \expval*{\hat{O}} \expval*{\hat{O} \hat{L}^\dagger \hat{L}}\}.
}
Using the cumulant expansion to two-point correlations, each term simplifies as
\eq{
    &\expval*{[\hat{A}, \hat{O}^2]} - 2 \expval*{\hat{O}} \expval*{[\hat{A}, \hat{O}]} = \expval*{\hat{A}} \expval*{\hat{O}^2} + 2 \expval*{\hat{O}} \expval*{\hat{A} \hat{O}} - \expval*{A} \expval*{\hat{O}^2} - 2 \expval*{O} \expval*{\hat{O} \hat{A}} - 2 \expval*{\hat{O}} \expval*{[\hat{A}, \hat{O}]} = 0, \\
    &2 \expval*{\hat{L}^\dagger \hat{O}^2 \hat{L}} - \expval*{\hat{L}^\dagger \hat{L} \hat{O}^2} - \expval*{\hat{O}^2 \hat{L}^\dagger \hat{L}} = \expval*{\hat{L}^\dagger \hat{O}} \expval*{[\hat{O}, \hat{L}]} + \expval*{\hat{O} \hat{L}} \expval*{[\hat{L}^\dagger, \hat{O}]}, \\
    &2 \expval*{\hat{O}} \expval*{\hat{L}^\dagger \hat{O} \hat{L}} - \expval*{\hat{O}} \expval*{\hat{L}^\dagger \hat{L} \hat{O}} - \expval*{\hat{O}} \expval*{\hat{O} \hat{L}^\dagger \hat{L}} = \expval*{\hat{L}^\dagger} \expval*{\hat{O}} \expval*{[\hat{O}, \hat{L}]} + \expval*{\hat{L}} \expval*{\hat{O}} \expval*{[\hat{L}^\dagger, \hat{O}]}.
}
The numerator thus becomes
\eq{
    \Delta \hat{O}_{\text{H}}^2(t) = \expval*{\Delta \hat{O}^2} + 2 \gamma t \{\expval*{\hat{L}^\dagger \hat{O}}_{\text{c}} \expval*{[\hat{O}, \hat{L}]} + \expval*{\hat{O} \hat{L}}_{\text{c}} \expval*{[\hat{L}^\dagger, \hat{O}]}\},
}
which is Eq.~\eqref{eq:uni_nume}.

\section{Derivations for Eqs.~\eqref{eq:nonuni_deno} and~\eqref{eq:nonuni_nume}}
\label{sec:app_deri_nonuni}

Both $\hat{O}_{\text{I}}(t)$ and $\hat{A}_{\text{I}}(\tau)$ expand to first order in $\Omega$ as
\eq{
    \hat{O}_\text{I}(t) &= \hat{O} + i \Omega t [\hat{A}, \hat{O}], \\
    \hat{L}_\text{I}(t) &= \hat{L} + i \Omega t [\hat{A}, \hat{L}].
}
When the initial state is an eigenstate of $\hat{O}$, applying the cumulant expansion to three-point correlations expands the denominator to first order in $\Omega$ as
\eq{
    &\int_0^t [ 2 \expval*{\hat{L}_\text{I}^\dagger(\tau) \hat{O}_\text{I}(t) \hat{L}_\text{I}(\tau)} - \expval*{\hat{L}_\text{I}^\dagger(\tau) \hat{L}_\text{I}(\tau) \hat{O}_\text{I}(t)} - \expval*{\hat{O}_\text{I}(t) \hat{L}_\text{I}^\dagger(\tau) \hat{L}_\text{I}(\tau)}] \dd \tau \\
    &= t [2 \expval*{\hat{L}^\dagger \hat{O} \hat{L}} - \expval*{\hat{L}^\dagger \hat{L} \hat{O}} - \expval*{\hat{O} \hat{L}^\dagger \hat{L}}] \\
    &\quad + i \Omega t^2 [ \expval*{\hat{A} \hat{L}^\dagger \hat{O} \hat{L}} + \expval*{\hat{L}^\dagger \hat{A} \hat{O} \hat{L}} - \expval*{\hat{L}^\dagger \hat{O} \hat{A} \hat{L}} - \expval*{\hat{L}^\dagger \hat{O} \hat{L} \hat{A}} + \expval{\hat{L}^\dagger \hat{L} \hat{O} \hat{A}} - \expval*{\hat{A} \hat{O} \hat{L}^\dagger \hat{L}} ] \\
    &= t [2 \expval*{\hat{L}^\dagger \hat{O} \hat{L}} - \expval*{\hat{L}^\dagger \hat{L} \hat{O}} - \expval*{\hat{O} \hat{L}^\dagger \hat{L}}] \\
    &\quad + i \Omega t^2 \expval*{\hat{L}^\dagger} ( \expval*{\hat{A} \hat{O} \hat{L}} + \expval*{\hat{L} \hat{O} \hat{A}} - \expval*{\hat{O} \hat{A} \hat{L}} - \expval*{\hat{O} \hat{L} \hat{A}} ) + i \Omega t^2 \expval*{\hat{L}} ( \expval*{\hat{A} \hat{L}^\dagger \hat{O}} + \expval*{\hat{L}^\dagger \hat{A} \hat{O}} - \expval*{\hat{L}^\dagger \hat{O} \hat{A}} - \expval*{\hat{A} \hat{O} \hat{L}^\dagger} ) \\
    &= t [2 \expval*{\hat{L}^\dagger \hat{O} \hat{L}} - \expval*{\hat{L}^\dagger \hat{L} \hat{O}} - \expval*{\hat{O} \hat{L}^\dagger \hat{L}}] \\
    &\quad + i \Omega t^2 \expval*{\hat{L}^\dagger} (\expval*{[\hat{A}, \hat{O}] \hat{L}} + \expval*{\hat{L} [\hat{O}, \hat{A}]}) + i \Omega t^2 \expval*{\hat{L}} (\expval*{[\hat{O}, \hat{A}] \hat{L}^\dagger} + \expval*{\hat{L}^\dagger [\hat{A}, \hat{O}]}).
}

The numerator $\Delta \hat{O}_{\text{H}}^2(t)$ expands to first order in $\gamma$ and $\Omega$ as
\eq{
    \Delta \hat{O}_{\text{H}}^2(t) &= [\expval*{\hat{O}^2} - \expval*{\hat{O}}^2] + i \Omega t \{\expval*{[\hat{A}, \hat{O}^2]} - 2 \expval*{\hat{O}} \expval*{[\hat{A}, \hat{O}]}\} \\
    &\quad + \gamma t \{2 \expval*{\hat{L}^\dagger \hat{O}^2 \hat{L}} - \expval*{\hat{L}^\dagger \hat{L} \hat{O}^2} - \expval*{\hat{O}^2 \hat{L}^\dagger \hat{L}}\} - 2 \gamma t \{2 \expval*{\hat{O}} \expval*{\hat{L}^\dagger \hat{O} \hat{L}} - \expval*{\hat{O}} \expval*{\hat{L}^\dagger \hat{L} \hat{O}} - \expval*{\hat{O}} \expval*{\hat{O} \hat{L}^\dagger \hat{L}}\} \\
    &= 2 \gamma t \expval*{[\hat{L}^\dagger, \hat{O}][\hat{O}, \hat{L}]}.
}

\section{Unitary encoding under dephasing noise}
\label{sec:app_uni_dephasing}

\begin{figure}[t!]
    \centering
    \includegraphics[width=8.5cm]{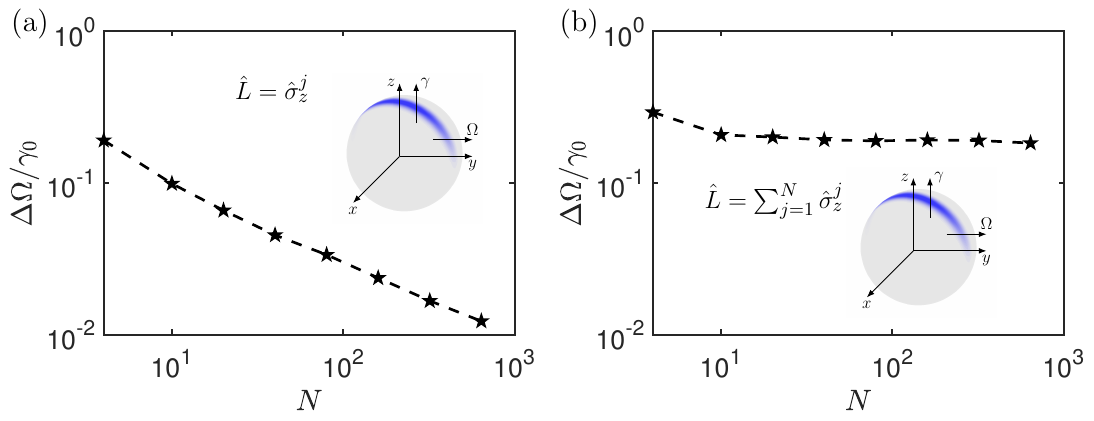}
    \caption{Precision scaling at the optimal encoding time for various $N$ under single-particle dephasing (a) and collective dephasing (b). The initial state is the optimal two-axis twisted state with $\expval*{\hat{J}_z} \propto N$ and $\expval*{\Delta \hat{J}_x} \propto 1$. The precision scaling degrades to $\Delta \Omega \propto 1 / \sqrt{N}$ under single-particle dephasing and $\Delta \Omega \propto 1$ under collective dephasing. The dissipation rate is $\gamma / \gamma_0 = 0.1$.}
    \label{fig:uni_dephasing_3}
\end{figure}

We also consider the case where $\hat{O}$ does not commute with $\hat{L}$, taking $\hat{O} = \hat{J}_x$ and $\hat{A} = \hat{J}_y$. For the optimal two-axis twisted state ($\expval*{\hat{J}_z} \propto N$, $\expval*{\Delta \hat{J}_x} \propto 1$), the precision at the optimal time degrades to $\Delta \Omega \propto 1 / \sqrt{N}$ for single-particle dephasing and $\Delta \Omega \propto 1$ for collective dephasing (Fig.~\ref{fig:uni_dephasing_3}), coinciding with the results in Fig.~\ref{fig:uni_dephasing_2}.

\section{Non-unitary encoding by dephasing processes}
\label{sec:app_nonuni_dephasing}

\begin{figure}[t!]
    \centering
    \includegraphics[width=8.5cm]{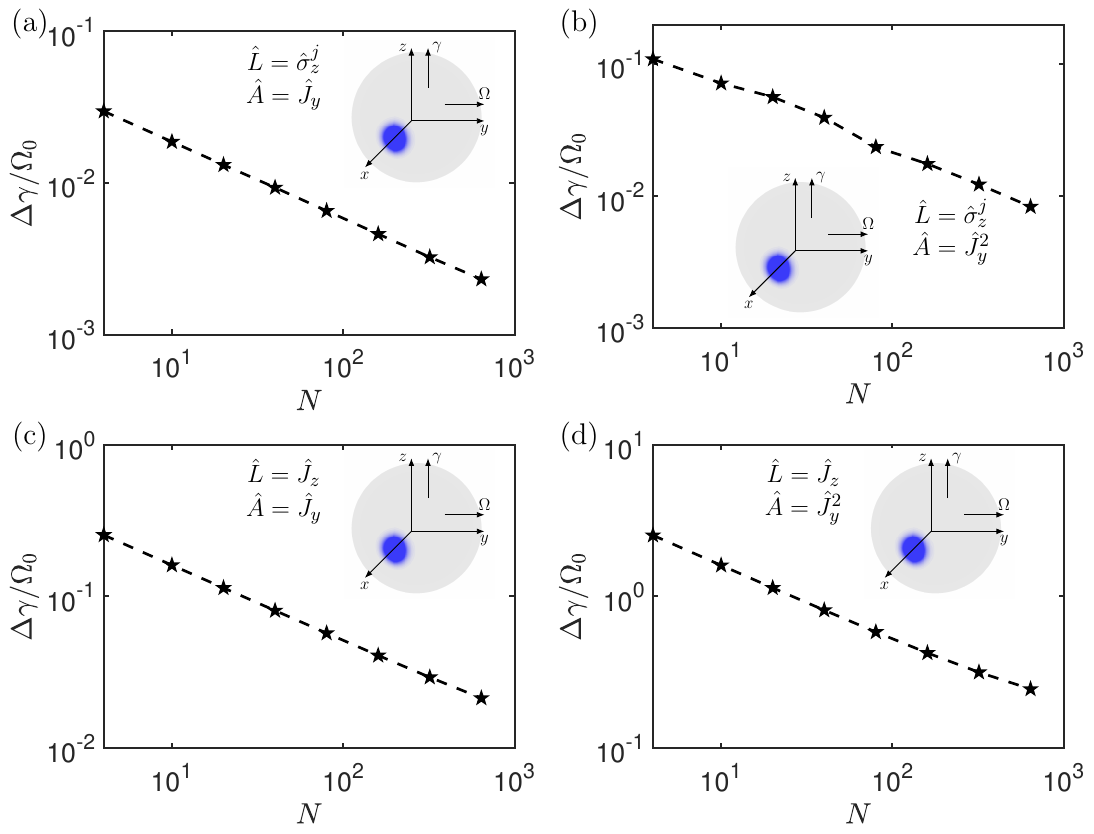}
    \caption{Precision scaling for various $N$ under non-unitary encoding $\hat{L} = \hat{\sigma}_z^j$ with intrinsic dynamics $\hat{A} = \hat{J}_y$ (a) and $\hat{A} = \hat{J}_y^2$ (b), and $\hat{L} = \hat{J}_z$ with intrinsic dynamics $\hat{A} = \hat{J}_y$ (c) and $\hat{A} = \hat{J}_y^2$ (d). The initial state is the eigenstate of $\hat{J}_x$ with eigenvalue $N$. The precision scales as $\Delta \gamma \propto 1 / \sqrt{N}$ in all cases. The intrinsic-dynamics strength is $\Omega / \Omega_0 = 0.1$. The energy and time units are $\Omega_0$ and $\Omega_0^{- 1}$, respectively.}
    \label{fig:nonuni_dephasing_2}
\end{figure}

\begin{figure}[t!]
    \centering
    \includegraphics[width=8.5cm]{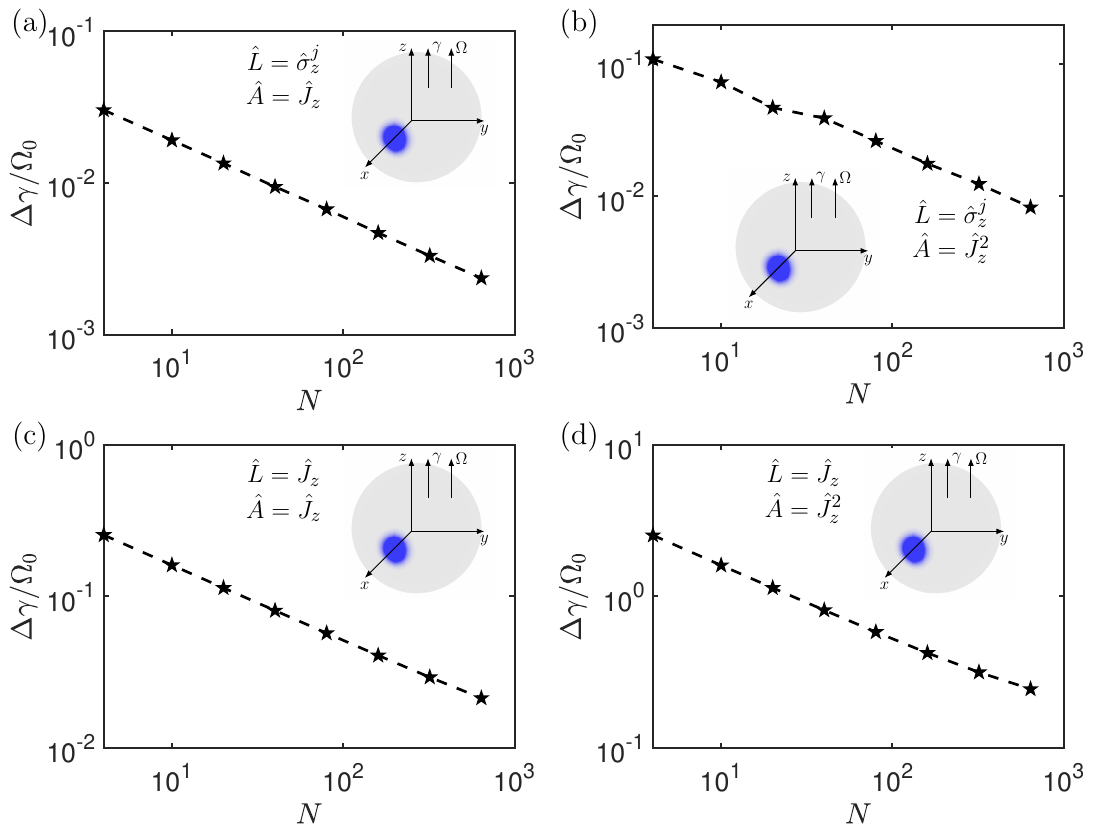}
    \caption{Precision scaling for various $N$ under non-unitary encoding $\hat{L} = \hat{\sigma}_z^j$ with intrinsic dynamics $\hat{A} = \hat{J}_z$ (a) and $\hat{A} = \hat{J}_z^2$ (b), and $\hat{L} = \hat{J}_z$ with intrinsic dynamics $\hat{A} = \hat{J}_z$ (c) and $\hat{A} = \hat{J}_z^2$ (d). The initial state is the eigenstate of $\hat{J}_x$ with eigenvalue $N$. The precision scales as $\Delta \gamma \propto 1 / \sqrt{N}$ in all cases. The intrinsic-dynamics strength is $\Omega / \Omega_0 = 0.1$. The energy and time units are $\Omega_0$ and $\Omega_0^{- 1}$, respectively.}
    \label{fig:nonuni_dephasing_3}
\end{figure}

Here we consider $[\hat{A}, \hat{O}] \neq 0$, with the same noise operators and observable as in Sec.~\ref{sec:nonuni_dephasing}. We take $\hat{A} = \hat{J}_y$ or $\hat{A} = \hat{J}_y^2$ in Fig.~\ref{fig:nonuni_dephasing_2}, and $\hat{A} = \hat{J}_z$ or $\hat{A} = \hat{J}_z^2$ in Fig.~\ref{fig:nonuni_dephasing_3}. The precision follows $\Delta \gamma \propto 1 / \sqrt{N}$ in all cases. These results confirm that the precision scaling is unaffected by intrinsic dynamics until the steady state is reached, even when $[\hat{A}, \hat{O}] \neq 0$.

\section{Non-unitary encoding by decay processes}
\label{sec:app_nonuni_decay}

\begin{figure}[t!]
    \centering
    \includegraphics[width=8.5cm]{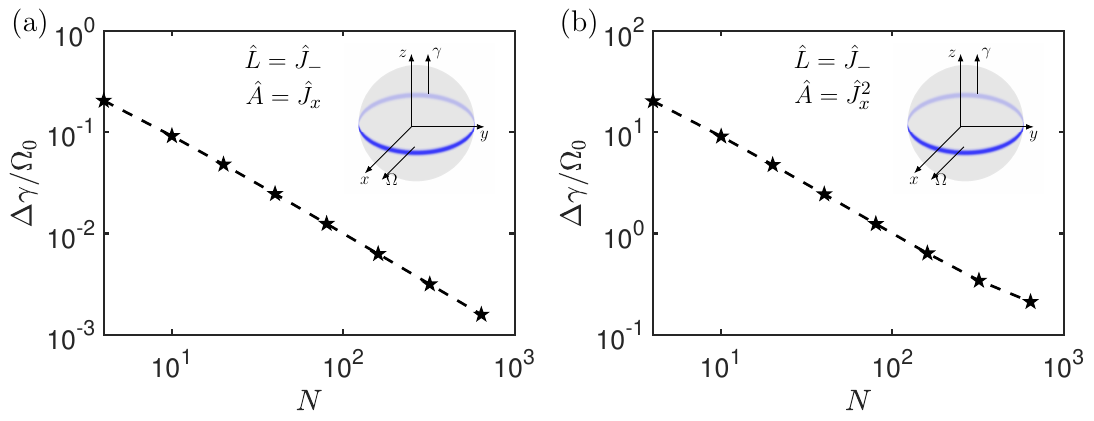}
    \caption{Precision scaling for various $N$ under non-unitary encoding $\hat{L} = \hat{J}_-$ with intrinsic dynamics $\hat{A} = \hat{J}_x$ (a) and $\hat{A} = \hat{J}_x^2$ (b), where the observable is $\hat{O} = \hat{J}_z$. The initial state is the Dicke state $\ket{N, 0}_z$ with eigenvalue $0$ of $\hat{J}_z$. The precision scales as $\Delta \gamma \propto 1 / N$ in both cases. The intrinsic-dynamics strength is $\Omega / \Omega_0 = 0.1$. The energy and time units are $\Omega_0$ and $\Omega_0^{- 1}$, respectively.}
    \label{fig:nonuni_decay_2}
\end{figure}

\begin{figure}[t!]
    \centering
    \includegraphics[width=8.5cm]{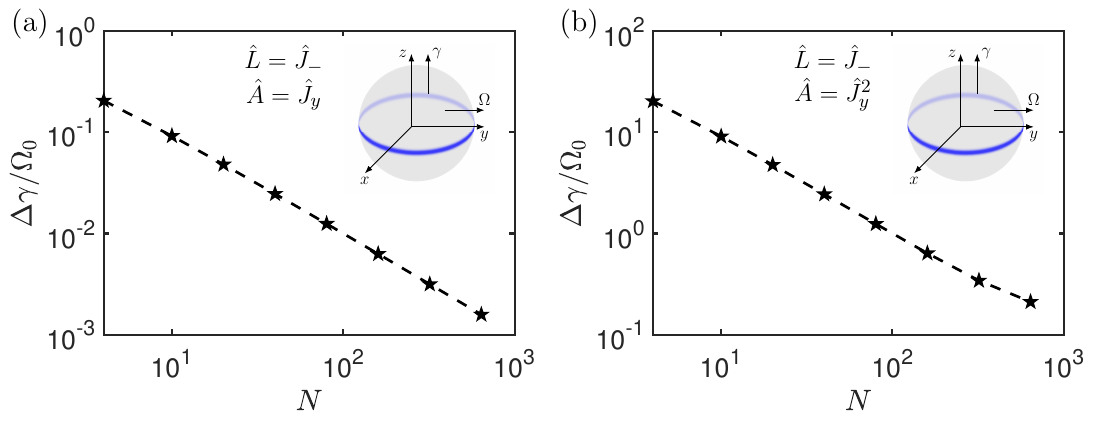}
    \caption{Precision scaling for various $N$ under non-unitary encoding $\hat{L} = \hat{J}_-$ with intrinsic dynamics $\hat{A} = \hat{J}_y$ (a) and $\hat{A} = \hat{J}_y^2$ (b), where the observable is $\hat{O} = \hat{J}_z$. The initial state is the Dicke state $\ket{N, 0}_z$ with eigenvalue $0$ of $\hat{J}_z$. The precision scales as $\Delta \gamma \propto 1 / N$ in both cases. The intrinsic-dynamics strength is $\Omega / \Omega_0 = 0.1$. The energy and time units are $\Omega_0$ and $\Omega_0^{- 1}$, respectively.}
    \label{fig:nonuni_decay_3}
\end{figure}

We also consider $[\hat{A}, \hat{O}] \neq 0$, taking $\hat{A} = \hat{J}_x$ or $\hat{A} = \hat{J}_x^2$ in Fig.~\ref{fig:nonuni_decay_2}, and $\hat{A} = \hat{J}_y$ or $\hat{A} = \hat{J}_y^2$ in Fig.~\ref{fig:nonuni_decay_3}. The precision follows $\Delta \gamma \propto 1 / N$ in all cases. These results confirm that the precision scaling is unaffected by intrinsic dynamics until the steady state is reached, even when $[\hat{A}, \hat{O}] \neq 0$.

\section{Cumulant expansions}
\label{sec:app_cumu}

The cumulant expansion expresses expectation values of operator products in terms of lower-order correlations. For operators $\hat{X}_1, \hat{X}_2, \hat{X}_3$, and $\hat{X}_4$ with non-zero mean values, it reads 
\eq{
    \expval*{\hat{X}_1 \hat{X}_2 \hat{X}_3} &= \expval*{\hat{X}_1 \hat{X}_2} \expval*{\hat{X}_3} + \expval*{\hat{X}_1 \hat{X}_3} \expval*{\hat{X}_2} + \expval*{\hat{X}_2 \hat{X}_3} \expval*{\hat{X}_1} - 2 \expval*{\hat{X}_1} \expval*{\hat{X}_2} \expval*{\hat{X}_3}, \\
    \expval*{\hat{X}_1 \hat{X}_2 \hat{X}_3 \hat{X}_4} &= \expval*{\hat{X}_1 \hat{X}_2 \hat{X}_3} \expval*{\hat{X}_4} + \expval*{\hat{X}_1 \hat{X}_2 \hat{X}_4} \expval*{\hat{X}_3} + \expval*{\hat{X}_1 \hat{X}_3 \hat{X}_4} \expval*{\hat{X}_2} + \expval*{\hat{X}_2 \hat{X}_3 \hat{X}_4} \expval*{\hat{X}_1} \\
    &\quad + \expval*{\hat{X}_1 \hat{X}_2} \expval*{\hat{X}_3 \hat{X}_4} + \expval*{\hat{X}_1 \hat{X}_3} \expval*{\hat{X}_2 \hat{X}_4} + \expval*{\hat{X}_1 \hat{X}_4} \expval*{\hat{X}_2 \hat{X}_3} \\
    &\quad - 2 (\expval*{\hat{X}_1 \hat{X}_2} \expval*{\hat{X}_3} \expval*{\hat{X}_4} + \expval*{\hat{X}_1 \hat{X}_3} \expval*{\hat{X}_2} \expval*{\hat{X}_4} + \expval*{\hat{X}_1 \hat{X}_4} \expval*{\hat{X}_2} \expval*{\hat{X}_3})\\
    &\quad - 2 (\expval*{\hat{X}_2 \hat{X}_3} \expval*{\hat{X}_1} \expval*{\hat{X}_4} + \expval*{\hat{X}_2 \hat{X}_4} \expval*{\hat{X}_1} \expval*{\hat{X}_3} + \expval*{\hat{X}_3 \hat{X}_4} \expval*{\hat{X}_1} \expval*{\hat{X}_2}) \\
    &\quad + 6 \expval*{\hat{X}_1} \expval*{\hat{X}_2} \expval*{\hat{X}_3} \expval*{\hat{X}_4}.
}

}

\normalem
% \bibliography{Precision_Open_Systems.bib}

\begin{thebibliography}{51}
\bibitem [{not()}]{note1}%
  \BibitemOpen
  \href@noop {} {}\bibinfo {note} {The error propagation formula is $\Delta \phi = \frac{\Delta \hat{O}_{\text{H}}(t)}{\sqrt{M} |\partial \expval*{\hat{O}(t)} / \partial \phi|}$}\BibitemShut {NoStop}%
\end{thebibliography}

\end{document}